\documentstyle[epsfig]{aipproc}

\newcommand{\nn}{\nonumber}
\renewcommand{\d}{{\rm d}}
\renewcommand{\u}{{\rm u}}

\newcommand{\ubar}{\overline{\rm u}}
\newcommand{\dbar}{\overline{\rm d}}
\newcommand{\sbar}{\overline{\rm s}}

\begin{document}
\title{Wide Angle Compton Scattering}

\author{Rainer Jakob}
\address{Fachbereich Physik, Universit\"at Wuppertal, D-42097 Wuppertal,
   Germany}

\maketitle

\begin{abstract}
We present the handbag contribution to Wide Angle Compton Scattering (WACS)
at moderately large momentum transfer obtained with a proton distribution 
amplitude close to the asymptotic form. In comparison it is found to be 
significantly larger than results from the hard scattering (pQCD) approach.
\end{abstract}

\section*{Introduction}
Compton scattering off nucleons provides us with valuable information on the
nucleon structure. In general, the Compton process involves the relativistic 
propagation of an excited, composite system, a bound state, which is
extremely difficult to describe. Fortunately, in special kinematic situations 
the description of the process becomes much simpler.
In this contribution we focus on Compton scattering off protons 
with large momentum transfer and small (or zero) photon 
virtuality, i.e.\ Compton scattering at wide angles (WACS). Here the 
process receives its main contribution from the lowest Fock state of the 
nucleon, i.e. a three quark configuration. The propagation of hard gluons 
and quarks is described perturbatively.
We argue that in the region of large, but not yet asymptotically 
large, momentum transfer the process is dominated by the handbag 
contribution involving new Compton form factors defined from skewed 
parton distributions. We compare the handbag contribution calculated 
from the overlap of soft wave fuctions~\cite{Diehl:1999kh,Diehl:1999tr} 
with the results obtained in the hard scattering 
approach~\cite{Brooks:2000nb}.
 
\section*{Hard vs. soft scattering mechanism}

\begin{figure}[ht!] 
\begin{center} 
\epsfig{file=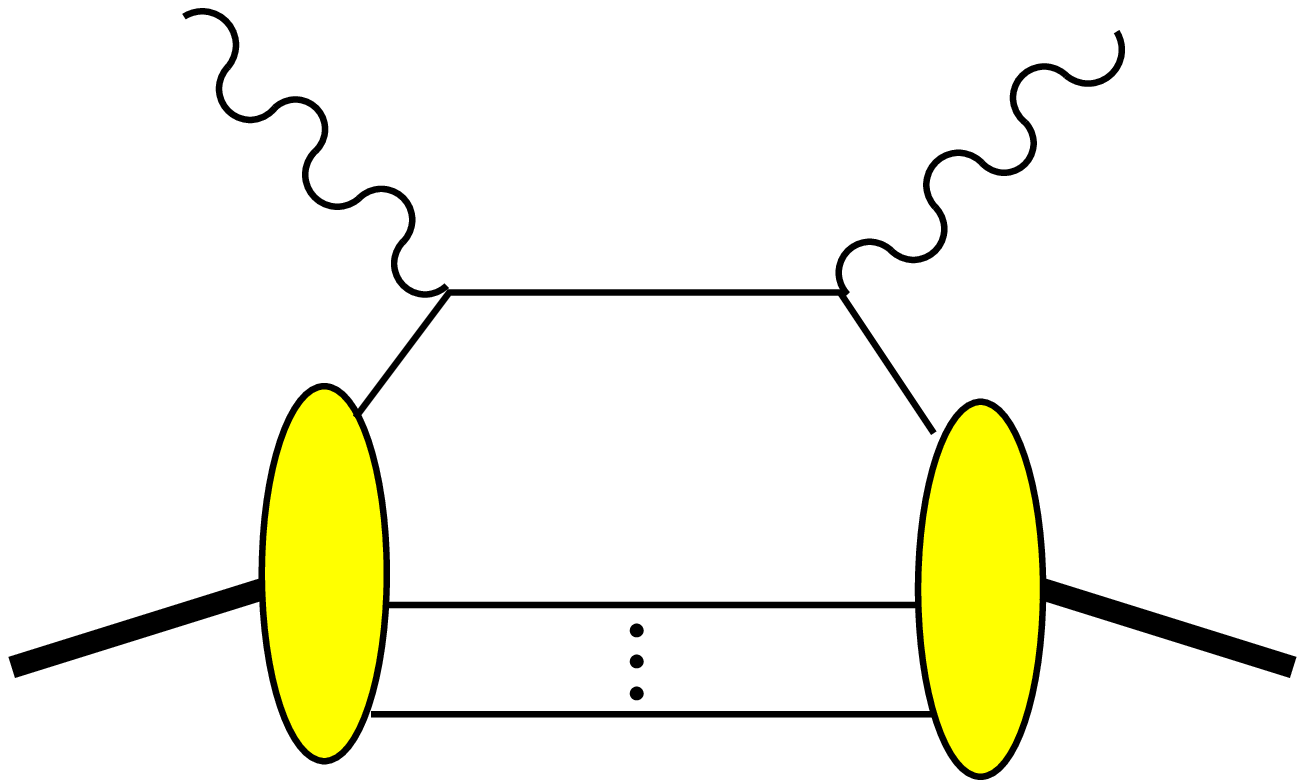,width=0.17\textwidth}
\raisebox{2ex}{$\;+\;$}
\epsfig{file=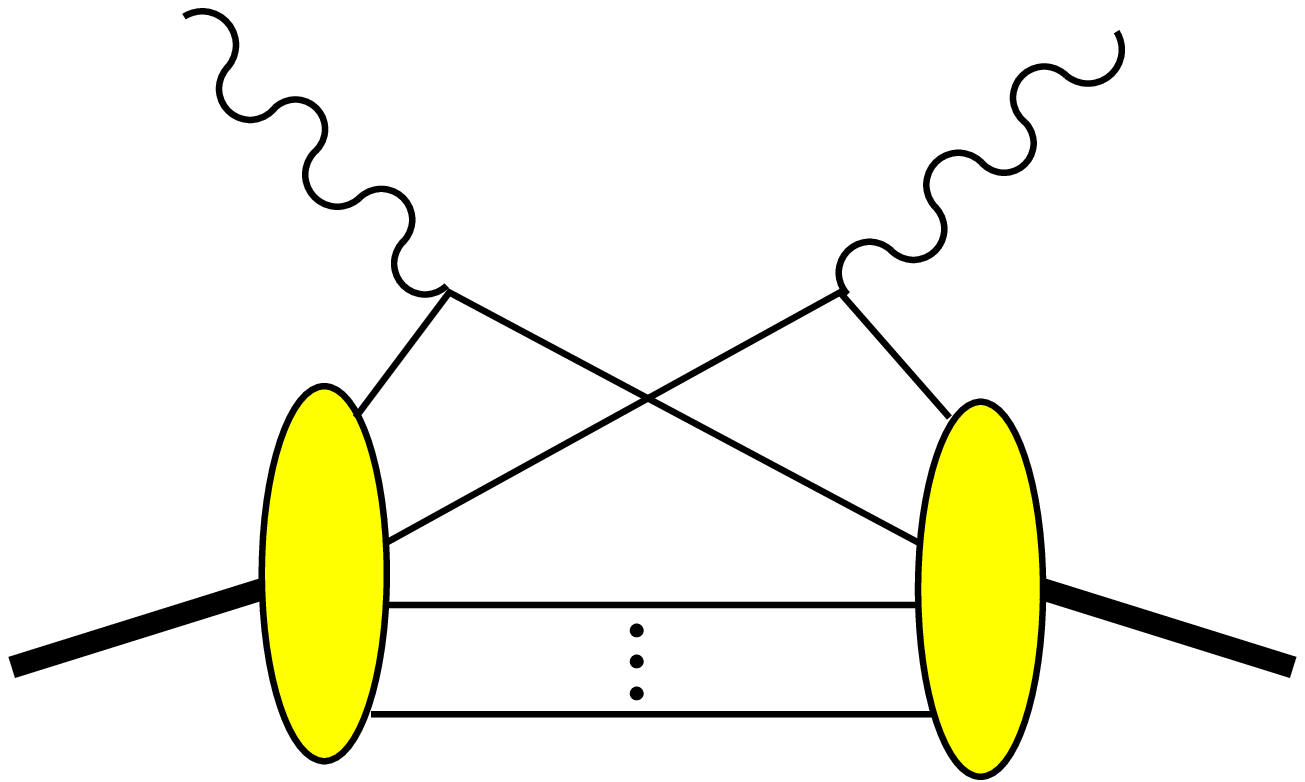,width=0.17\textwidth}
\raisebox{2ex}{$\;+\ldots+\;$}
\epsfig{file=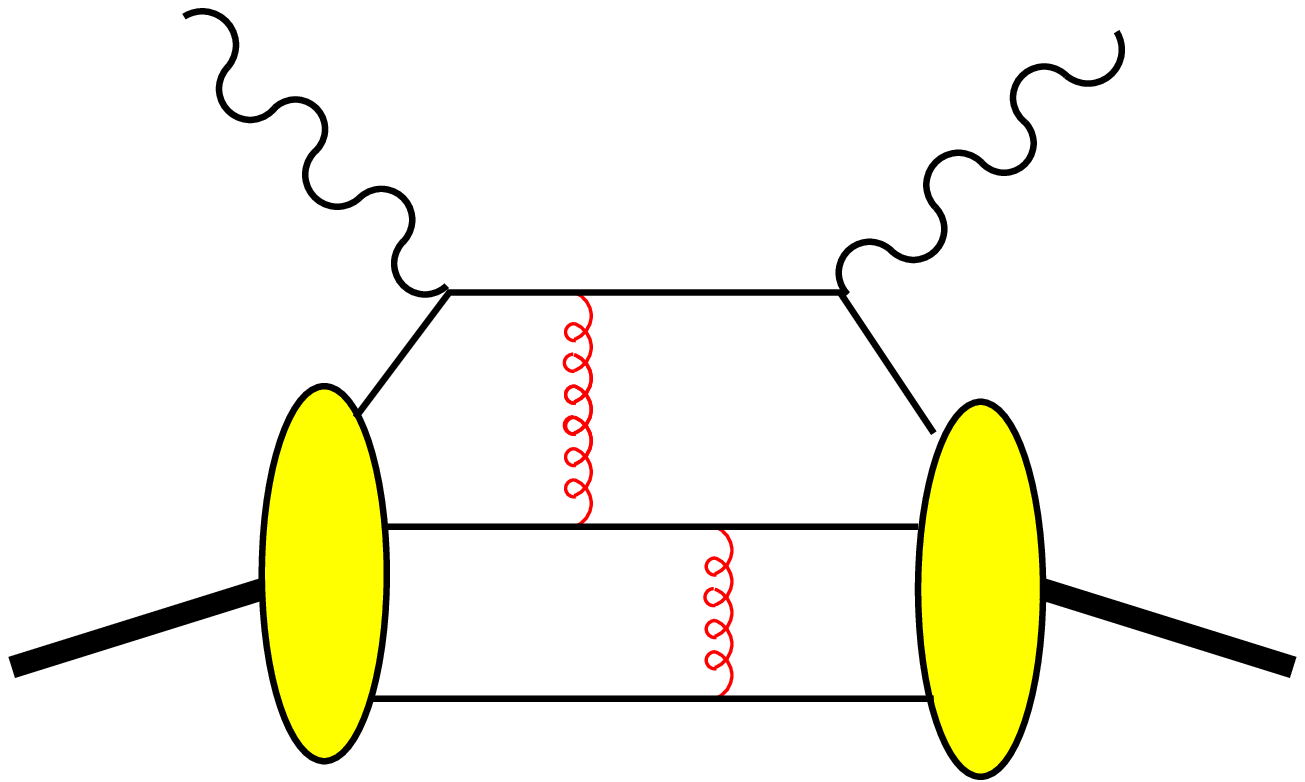,width=0.17\textwidth}
\raisebox{2ex}{$\;+\;$}
\epsfig{file=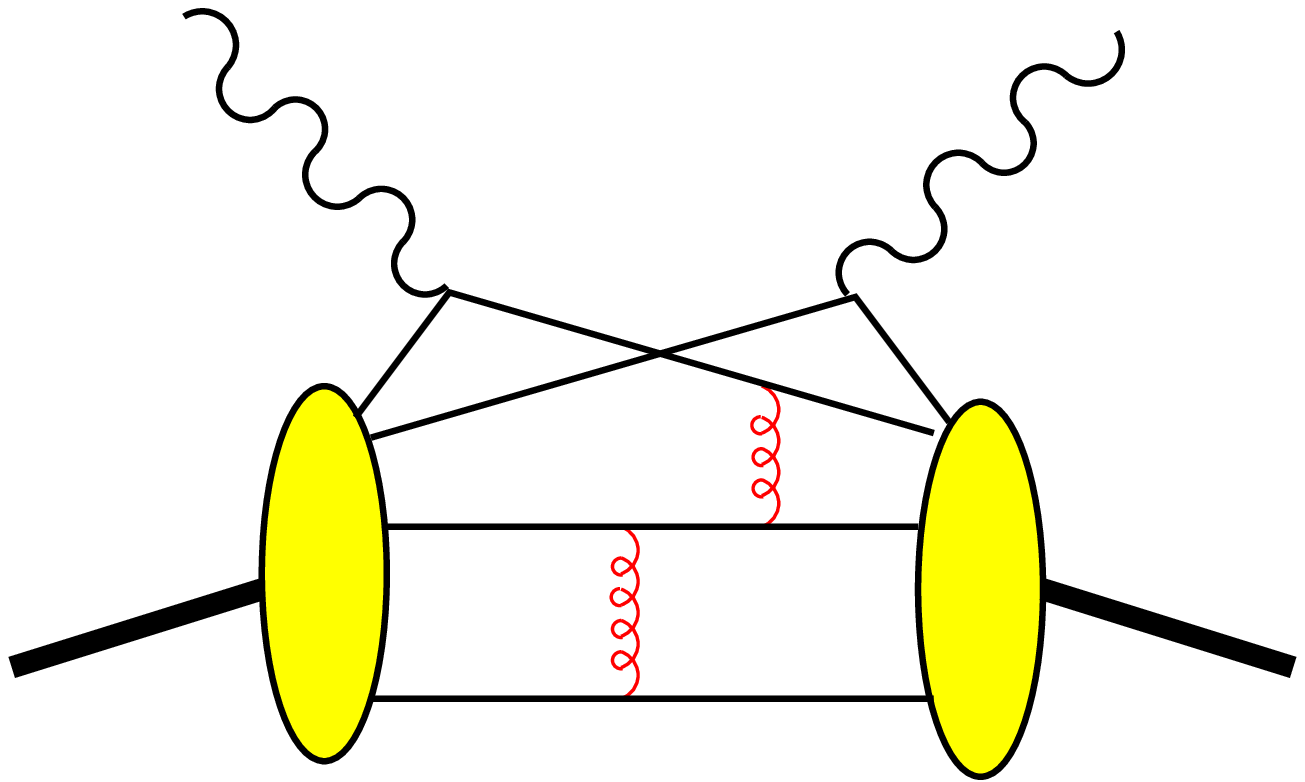,width=0.17\textwidth}
\raisebox{2ex}{$\;+\ldots$}
\end{center} 
\vspace*{-2mm}
\hspace*{0.08\textwidth}{\bf a}
\hspace*{0.2\textwidth}{\bf b}
\hspace*{0.26\textwidth}{\bf c}
\hspace*{0.22\textwidth}{\bf d}\\[1mm]
\caption{Some of the diagrams contributing to WACS. The dots between {\bf b} 
  and {\bf c} indicate an intermediate class of diagrams with one exchanged 
  hard gluon, whereas the dots behind {\bf d} stand for diagrams with a 
  higher number of gluons ($\alpha_s$ corrections), and diagrams from 
  higher Fock states (power corrections).}
\label{F:rjakob:1}
\end{figure}

It is generally 
accepted that the hard scattering mechanism in the context of perturbative 
QCD~\cite{Lepage:1980fj} provides the correct description 
for processes at asymptotically large 
momentum transfer. In this picture a hard scattering amplitude describes 
the redistribution of the transfered momentum among partons via hard gluon
exchange, and the non-perturbative part is given as distribution amplitudes
(DAs), i.e. light-cone wave functions (LCWFs) integrated over transverse
momenta. There is a minimal number of hard gluons required to connect all 
parton lines, the lowest Fock state gives the dominant contribution. In 
Fig.~\ref{F:rjakob:1} the diagrams {\bf c} and {\bf d} (and diagrams 
obtainable from them by permutations) stand for the hard scattering picture. 

For exclusive processes at moderately large momentum transfer, on the other
hand, there is a longstanding debate on the question which is the dominant 
reaction mechanism.
The diagrams {\bf a} and {\bf b} in Fig.~\ref{F:rjakob:1} stand for the 
Feynman mechanism, to be calculated from the direct overlap of soft 
LCWFs. No redistribution of the momentum transfer is necessary, if the
momentum fraction of the active parton line is large. The wave functions
suppress such asymmetric configurations; asymptotically these diagrams
will be power corrections to the hard scattering ones. At intermediate
momentum transfer, however, the relative importance cannot be 
judged {\em a priori}. Explicite numerical calculations are necessary.

\section*{WACS in the hard scattering approach}

For a comparison we will quote the results from a recent leading order 
calculation of real WACS off protons in the hard scattering 
approach~\cite{Brooks:2000nb} superseding (and partly correcting) earlier 
calculations~\cite{Kronfeld:1991kp,Vanderhaeghen:1997}.

\begin{figure}[h!] 
\begin{center} 
\epsfig{file=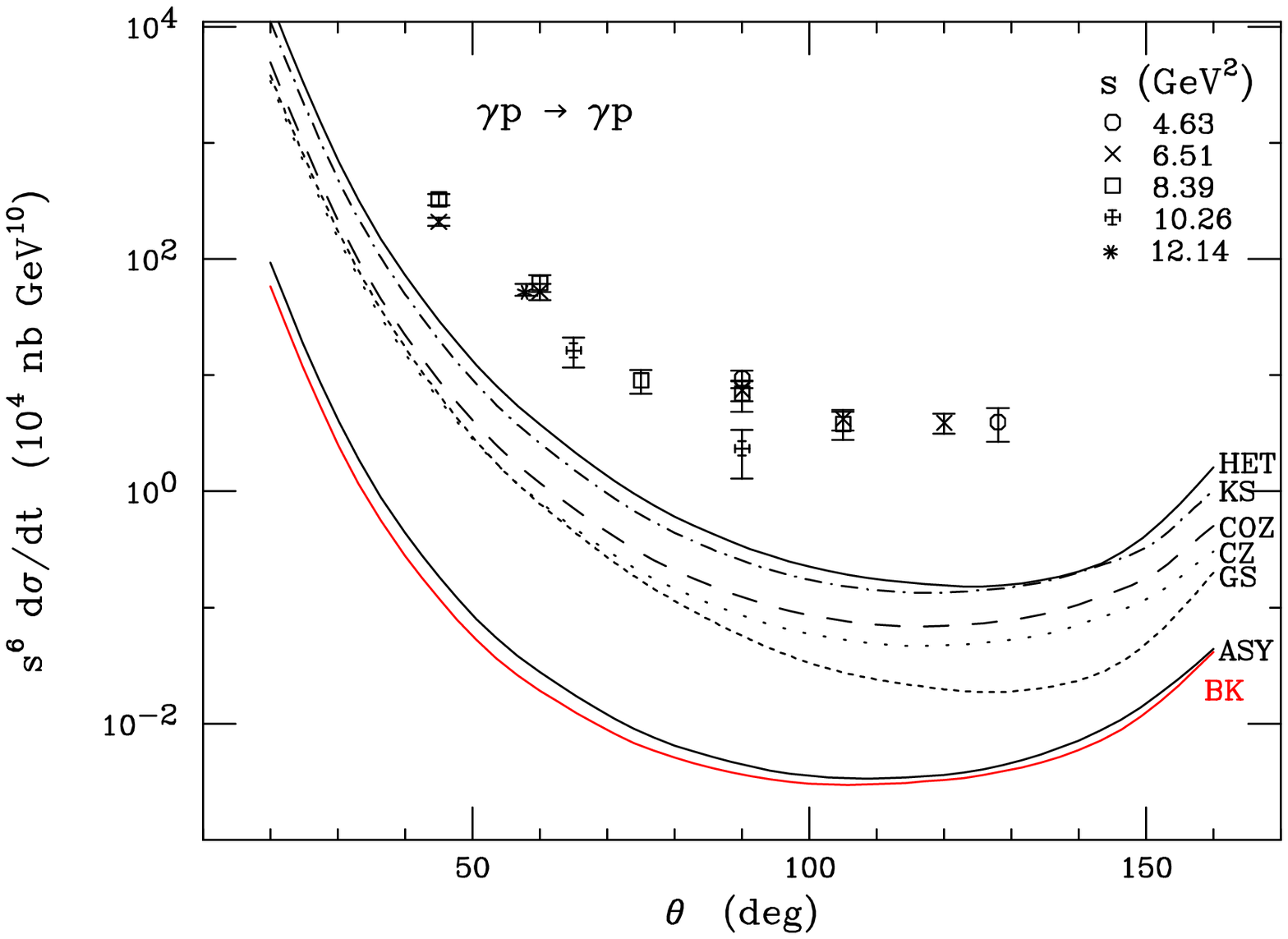,width=0.46\textwidth}
\qquad
\epsfig{file=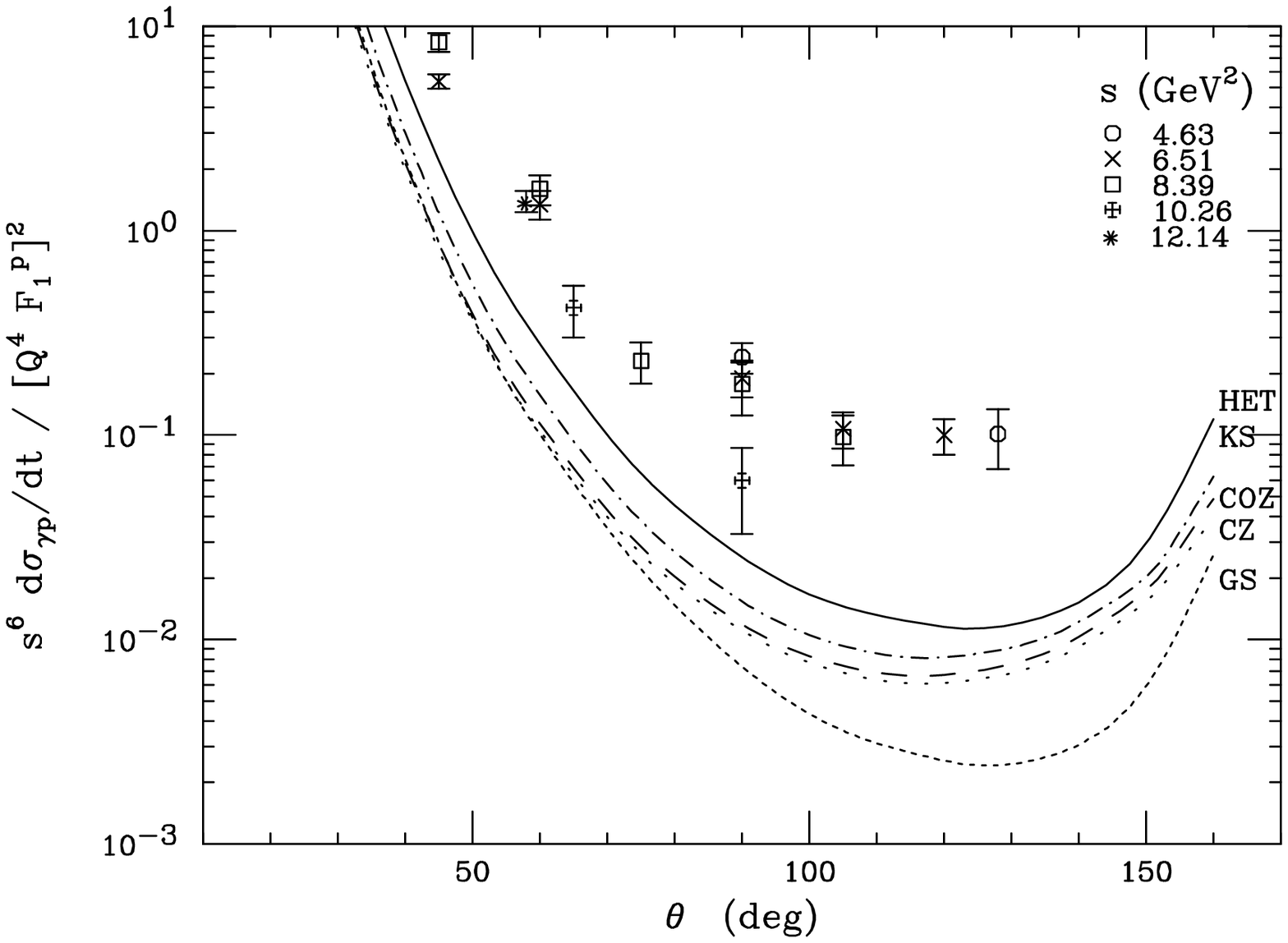,width=0.46\textwidth}
\end{center} 
\caption[fig2]{Left: The hard scattering contribution to the unpolarised 
  differential cross section for WACS obtained with different DAs compared 
  to the data~\cite{Shupe:1979vg}.  
  Right: The same quantity scaled by the pQCD result for 
  $(Q^4\,F_1^p(Q^2))^2$. Figures reproduced from~\cite{Brooks:2000nb} with
  one additional curve (BK) provided by the same authors.}
\label{F:rjakob:2}
\end{figure}

The unpolarised cross section, scaled by $s^6$, obtained with different 
DAs is shown in Fig.~\ref{F:rjakob:2} in comparison to the data. To 
minimize the 
influence of choices for the $\alpha_s(\mu)$ argument, the same quantity
normalized to the factor $(Q^4\,F_1^p(Q^2))^2$, where $F_1^p$ is the Dirac 
form facor of the proton, is also shown. Uncertainties are
expected to cancel from this ratio to a large extent. 

Clearly, the results of the hard scattering (pQCD) contributions to the
unpolarised cross sections fall short to describe the available data, which
are admittedly at rather lowish momentum transfers. From the ratio shown on
the RHS the authors of\cite{Brooks:2000nb} conclude that {\em `\ldots it 
seems unlikely that the elastic proton form factor and the Compton 
scattering amplitudes are both described by pQCD at presently accessible 
energies'}.

These results suggest, that hard scattering is not the dominant reaction
mechanism at the intermediate large momentum transfer, a situation very
similar to the elastic nucleon (and pion) form factors. 

\section*{WACS  in the soft physics approach}

The contribution from the Feynman mechanism to WACS is described by a 
handbag diagram~\cite{Radyushkin:1998rt}. The diagram factorises in a 
hard photon-parton amplitude and a non-perturbative 
part~\cite{Diehl:1999kh}, the latter 
described by a skewed parton distribution at vanishing 'skewedness', which 
can be calculated from the overlap of LCWFs, as indicated
by diagram {\bf a} in Fig.~\ref{F:rjakob:1}. The cat-ears diagram {\bf b} 
was shown to be power suppresssed relativ to the handbag 
diagram~\cite{Diehl:1999kh}. 
The (unpolarised) differential cross section can be written
\begin{equation} 
\frac{\d\sigma}{\d t} \;=\; \frac{2\pi\alpha_{\it em}^2}{s^2} \,
                            \left[-\frac{u}{s} - \frac{s}{u}\right]
\left\{ \frac{1}{2} \, 
   \left(R_V^2(t) + R_A^2(t)\right)
          -\frac{us}{s^2+u^2}\, 
   \left(R_V^2(t) - R_A^2(t)\right)\right\}
\label{E:rjakob:xsec}
\end{equation} 
with new form factors~\cite{Radyushkin:1998rt} specific to Compton scattering 
depending on $-t$ only
\begin{eqnarray}
\lefteqn{ \sum_a e_a^2\, \int_0^1 \frac{\d x}{x} \, p^+
   \int {\d z^-\over 2\pi}\, e^{i\, x p^+ z^-}\;
     \langle p'|\,
     \overline\psi{}_{a}(0)\, \gamma^+\,\psi_{a}(z^-) - 
     \overline\psi{}_{a}(z^-)\, \gamma^+\,\psi_{a}(0) 
     \,| p\rangle}
\nn\\&&\hspace{26mm}
 = R_V(t)\, \bar{u}(p')\, \gamma^+ u(p)\, 
   + R_T(t)\, \frac{i}{2m}\, \bar{u}(p')\, \sigma^{+\nu}
                  \Delta_\nu \, u(p) \;.
\end{eqnarray}
$R_T$ being related to nucleon helicity flips is neglected in 
Eq.~(\ref{E:rjakob:xsec}). An analogous definition holds for $R_A$ involving 
the axialvector nucleon matrix element. 

\begin{figure}[h!] 
\begin{center} 
\epsfig{file=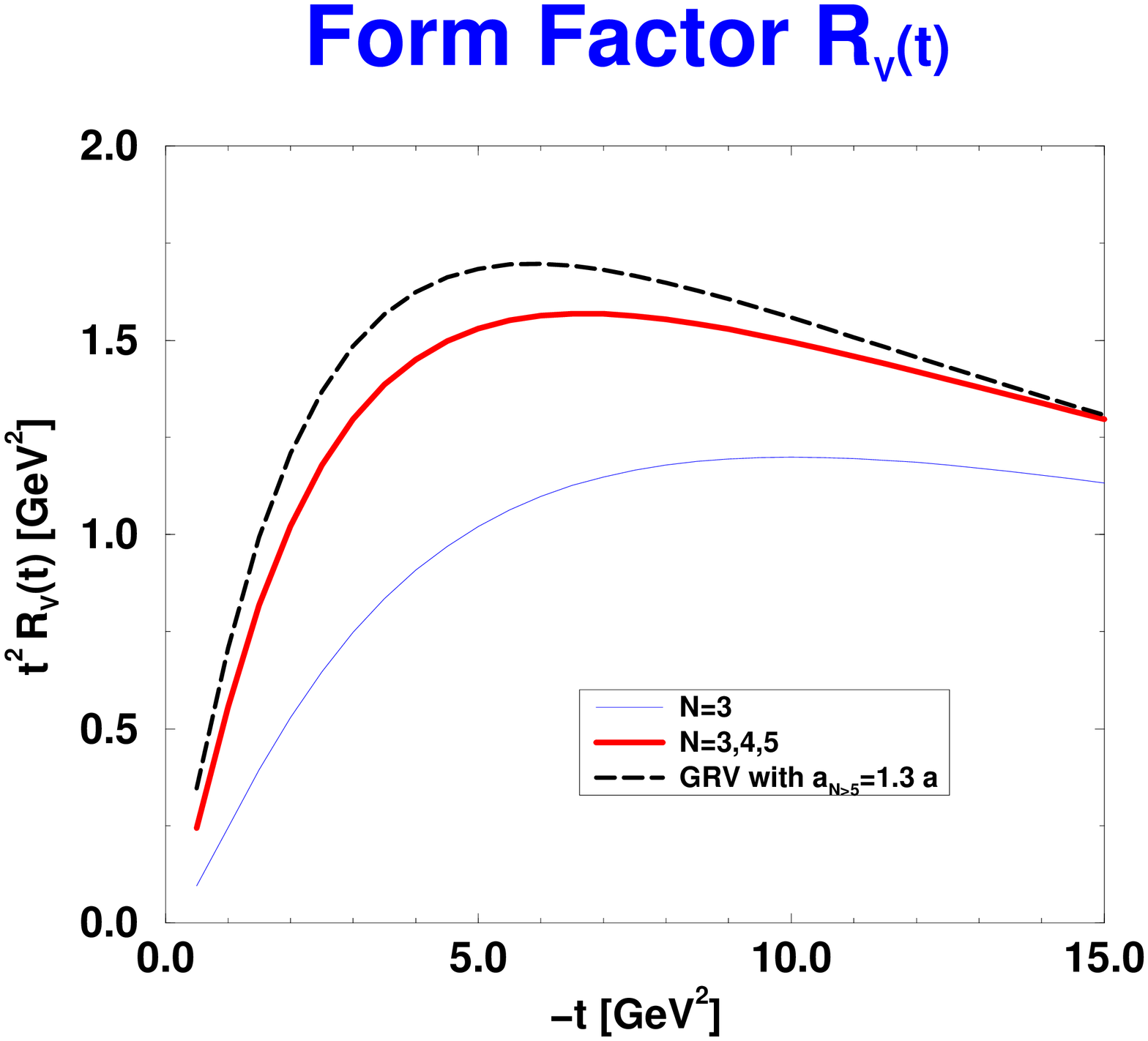,angle=-90,width=0.32\textwidth}
\qquad\quad
\epsfig{file=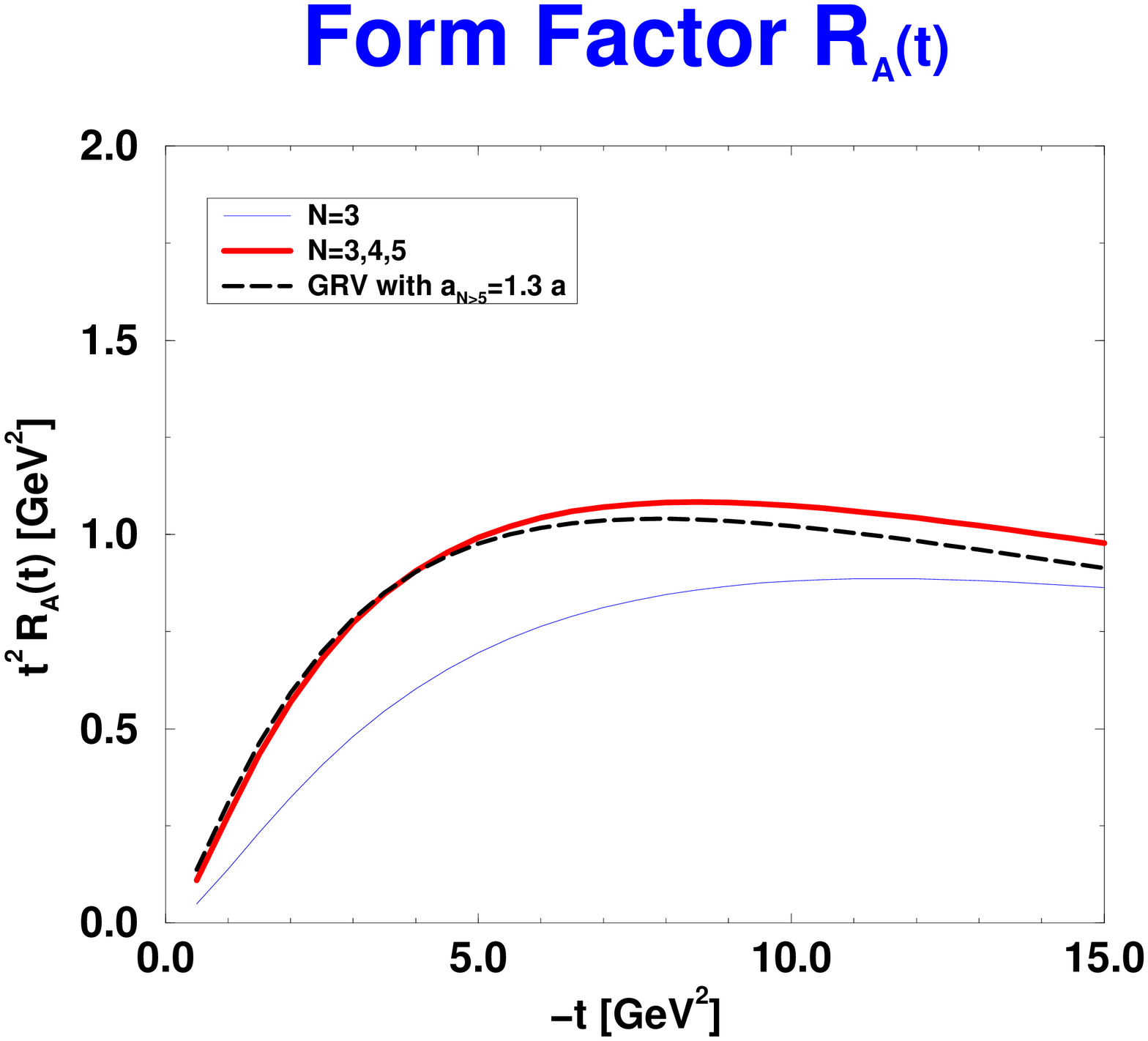,angle=-90,width=0.32\textwidth}
\end{center} 
\caption[fig3]{CS Form factors calculated from the valence Fock state only
(thin solid line), N=3,4,5 Fock states (thick solid line), and with an
additional estimate for higher Fock states(dashed line). Right: $R_V(t)$.
Left: $R_A(t)$.}
\label{F:rjakob:3}
\end{figure}

\begin{figure}[h!] 
\begin{center} 
\epsfig{file=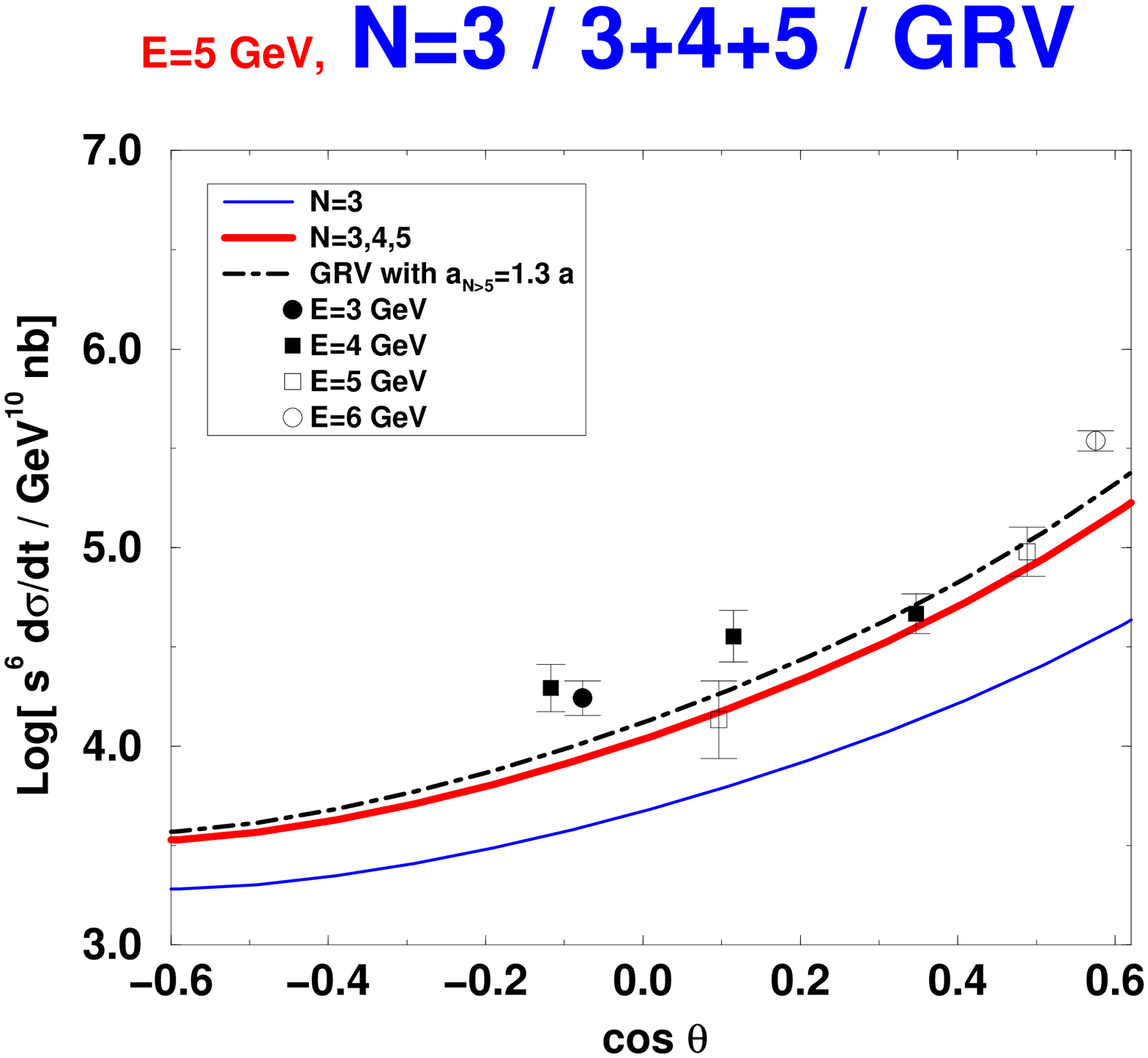,angle=-90,width=0.4\textwidth}
\qquad
\epsfig{file=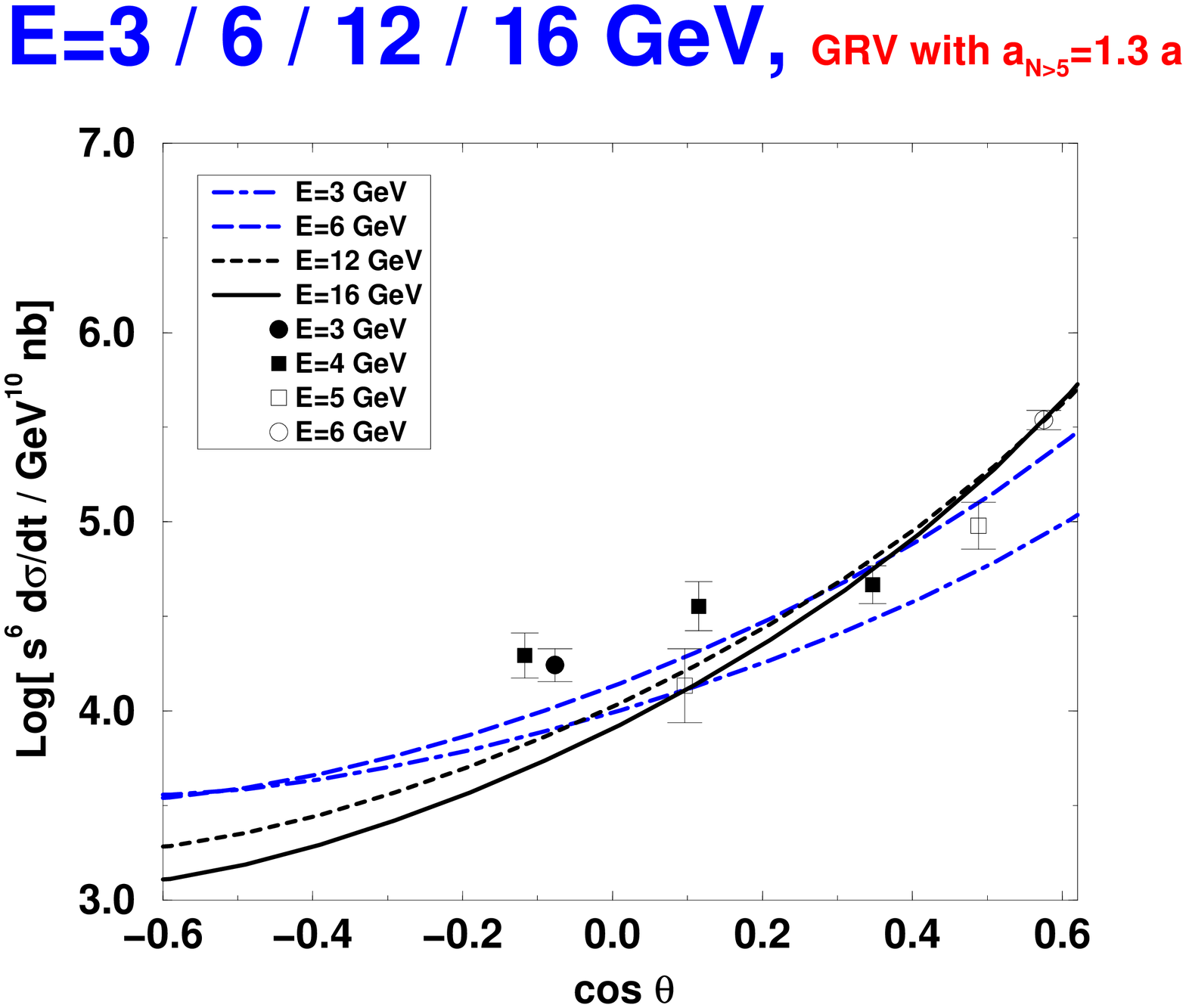,angle=-90,width=0.4\textwidth}
\end{center} 
\caption[fig4]{Cross sections for WACS obtained from the handbag diagram. 
  Data are from \cite{Shupe:1979vg}. 
  Left: for $E=5\,$GeV the contributions from the valence Fock  state, 
  the N=3,4,5 Fock states, and for all Fock states are compared. 
  Right: the cross sections for different photon energies obtained by the 
  `all Fock state estimate'.

Data from \cite{Shupe:1979vg}.}
\label{F:rjakob:4}
\end{figure}

Assuming a Gaussian model for transverse parton momenta
the form factors factorise in ordinary parton distribution functions (PDFs) 
and a $(t,x)$ dependent exponential for each $N$ parton Fock state 
separately
\begin{eqnarray} 
\lefteqn{
R_V^{(N)}(t) = \int_0^1 \frac{\d x}{x} 
       \exp{\left[ \frac{a^2_N \, t}{2} \, \frac{1-x}{x} \right]}}
\nn\\&&\times
   \left\{\, e_u^2 \; [ \u_v^{(N)}(x) +  2 \, \ubar^{(N)}(x)] 
         {}+ e_d^2 \; [ \d_v^{(N)}(x) +  2 \, \dbar{}^{(N)}(x)]
                      {}+ e_s^2 \; 2 \, \sbar{}^{(N)}(x)  
     \right\} \;,
\end{eqnarray}
and analogously for $R_V\to R_A$, $q(x)\to \Delta q(x)$.
With a phenomenologically based model for the $x$-dependence of the LCWF, 
the BK distribution amplitude~\cite{Bolz:1996sw} for the lowest Fock 
state, the form
factors $R_V$ and $R_A$ can be calculated. The results are shown in
Fig.~\ref{F:rjakob:3} together with estimates for the additional contributions
from the next higher Fock states (N=4,5), and an estimate for the the 
effect of all Fock states based on parametrizations for 
PDFs~\cite{Gluck:1998xa}. For details see Ref.~\cite{Diehl:1999kh}.
Inserting the form factors into Eq.~(\ref{E:rjakob:xsec}) the cross section
of WACS are obtained as displayed in Fig.~\ref{F:rjakob:4}. 
A comparison shows that the predictions for cross sections from the handbag 
diagram are much higher then the corresponding ones obtained in the hard
scattering picture and can fairly well describe the present 
data. Note that for a direct comparison a curve was added 
in Fig.~\ref{F:rjakob:2} (left) obtained within the hard scattering approach
in exactly the same way (same value for the parameter $f_N$) as the other 
curves in Fig.~\ref{F:rjakob:2}, but with the BK distribution amplitude as
input.  

Of particular interest is the initial state helicity correlation
\begin{equation} 
A_{\rm LL}\, \frac{{\rm d} \sigma}{{\rm d} t} \:=\:
\frac{1}{2} \;
   \left(
    \frac{{\rm d} \sigma(\mu=+1,\nu=+1/2)}{{\rm d} t}
   -\frac{{\rm d} \sigma(\mu=+1,\nu=-1/2)}{{\rm d} t} \right)
\end{equation} 
where $\mu,\nu$ are the helicities of the incoming photon and proton,
respectively. The prediction for this quantity from the handbag diagram is
distincively different from predictions obtained in the hard scattering
approach, or the diquark model~\cite{Kroll:1991kh}.

\begin{figure}[h!] 
\begin{center} 
\begin{minipage}{0.4\textwidth} 
\vspace*{-2ex}
\epsfig{file=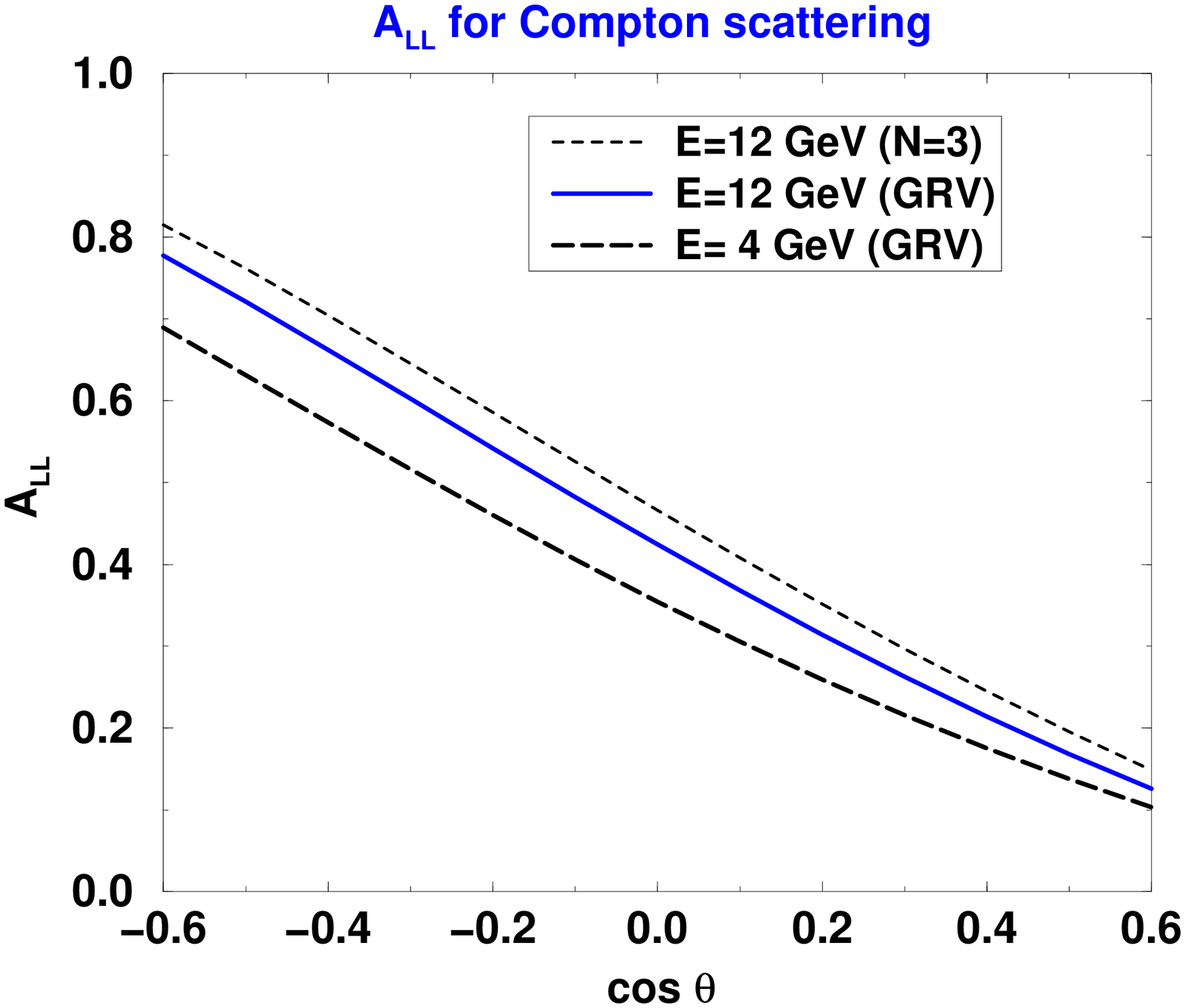,angle=-90,width=\textwidth}
\end{minipage} 
\qquad
\begin{minipage}{0.4\textwidth} 
\epsfig{file=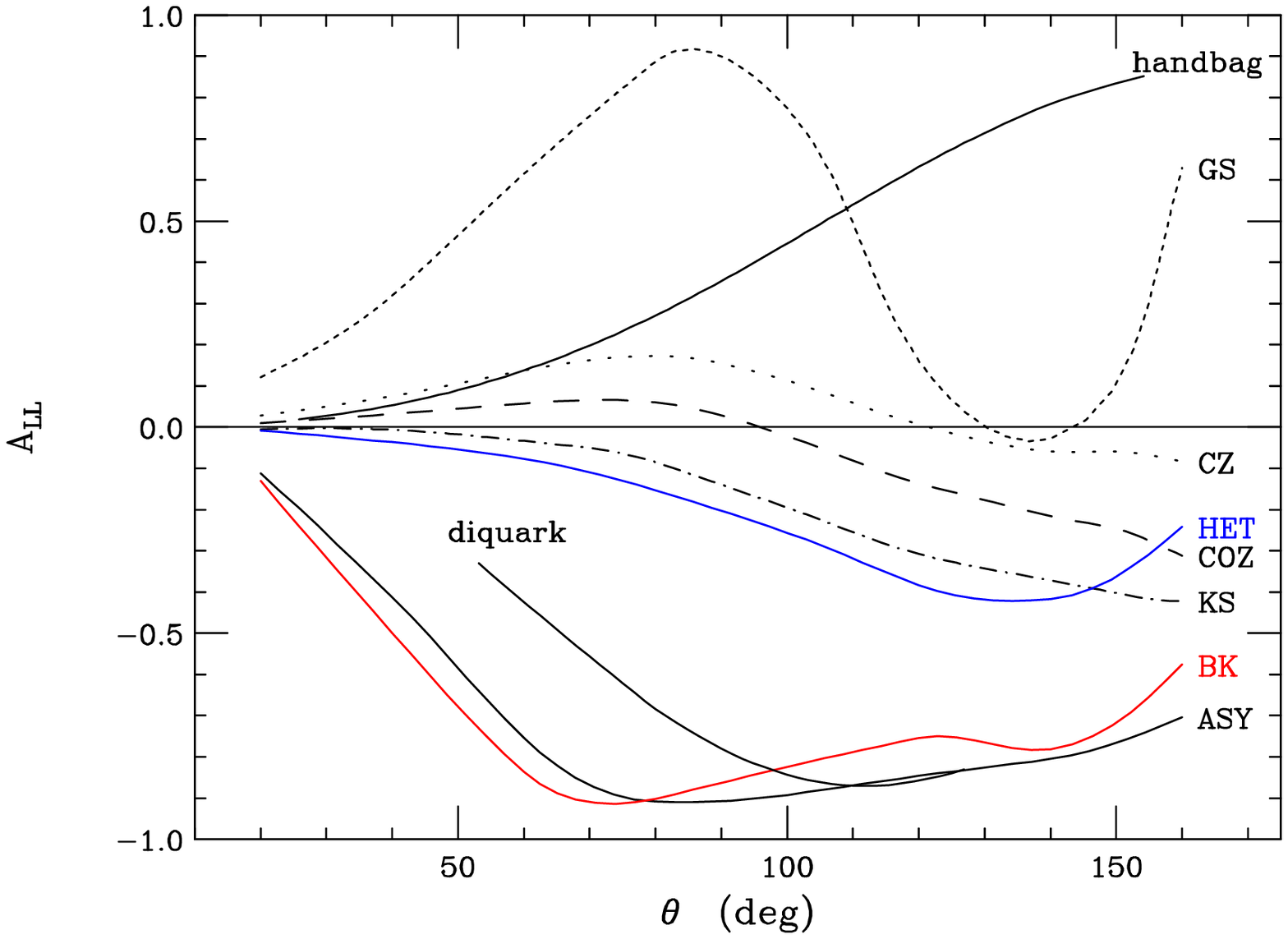,width=\textwidth}
\end{minipage} 
\end{center} 
\caption[fig2]{Initial state helicity correlation. Left: Predictions
of the handbag contribution for different energies. Right: Comparison
of different predictions at $E=4\,$GeV (taken from~\cite{Brooks:2000nb}).}
\label{F:rjakob:5}
\end{figure}


\vspace*{-3.4ex}
\section*{Acknowledgments}
This contribution is based on work done in collaboration with 
M.~Diehl, Th.~Feldmann, and P.~Kroll. I am grateful to T.~Brooks and 
L.~Dixon for the permission to reproduce their figures for comparison
and for providing the curve BK in Fig~\ref{F:rjakob:2}.

\end{document}